\documentclass[a4paper]{article}
\usepackage{amsfonts}    
\usepackage{amssymb}
\usepackage{latexsym}
\usepackage{graphicx}
\usepackage[usenames,dvipsnames]{color}

\newcommand{\pa}{\partial}
\newcommand{\ep}{\epsilon}

\newcommand{\ta}{\tau}
\newcommand{\ga}{\gamma}

\newcommand{\om}{\omega}
\newcommand{\de}{\delta}

\newcommand{\De}{\Delta}

\newcommand{\half}{\frac{1}{2}}
\newcommand{\rar}{\rightarrow}
\newcommand{\lrar}{\leftrightarrow}

\newcommand{\non}{\nonumber}

\begin{document}

\title{(Quasi)-exact-solvability on the sphere $S^n$}

\author{Willard Miller, Jr.\\
  School of Mathematics, University of Minnesota, \\
Minneapolis, Minnesota, U.S.A.\\
miller@ima.umn.edu\\[8pt]
and \\[8pt]
Alexander V Turbiner\\
Instituto de Ciencias Nucleares, UNAM, \\
 M\'exico DF 04510, Mexico\\
turbiner@nucleares.unam.mx}

\maketitle
\begin{abstract}
An Exactly-Solvable (ES) potential on  the sphere $S^n$ is reviewed and the related  Quasi-Exactly-Solvable (QES) potential is found and studied. Mapping the sphere to
a simplex it is found that the metric (of constant curvature) is in polynomial form, and both the ES and the QES potentials are rational functions. Their hidden algebra is
$gl_n$ in a finite-dimensional representation realized by first order differential
operators acting on $RP^n$.
It is shown that variables in the Schr\"odinger eigenvalue equation  can  be separated in
spherical coordinates and a number of the integrals of the second order exists assuring the complete integrability.
The QES system is completely-integrable for $n=2$ and non-maximally superintegrable for $n\ge 3$.
There is no separable coordinate system in which it is exactly solvable. We point out that by taking contractions of superintegrable systems, such as induced by  Wigner-In\"on\"u Lie algebra contractions, we can find other QES superintegrable systems, and we illustrate this by contracting our $S^n$ system to a QES non-maximal superintegrable system on Euclidean space $E^n$, an extension of the Smorodinsky-Winternitz potential.
\end{abstract}

\vskip 2cm



\section{Introduction} We are concerned with quantum mechanical eigenvalue problems
${\cal H}\Psi=E\Psi$ that can be solved exactly, algebraically, for the
eigenvalues and eigenvectors. Here, ${\cal H}=\De_g^{(n)} + V$ where $\De_g^{(n)}$ is the Laplace-Beltrami operator on an $n$-dimensional Riemannian or pseudo-Riemannian manifold and $V$ is a potential function on the manifold. There are four principal types
of systems that can be solved exactly:

\begin{itemize}
\item
   A system is {\bf exactly-solvable, (ES)} if there is the infinite flag of subspaces ${\cal P}_{j}$, $j=1,2, \cdots$, of the domain of ${\cal H}$
   such that $n_j=\dim {\cal P}_{j}\to \infty$ as $j \to \infty$ and
   ${\cal H}{\cal P}_{j}\subseteq {\cal P}_{j}\subset {\cal P}_{j+1}$ for each $j$. Note that for each subspace ${\cal P}_{j}$ the $n_j$ eigenvalues and eigenfunctions of ${\cal H}$ can be determined by algebraic means.

\item
   A system is {\bf quasi-exactly solvable, (QES)} if there is a single  subspace
   ${\cal P}_k$ of dimension $n_k>0$ such that
   ${\cal H} {\cal P}_{k}\subseteq {\cal P}_{k}$. In this case we can find  $n_k$ eigenvalues and eigenfunctions of ${\cal H}$ by algebraic means, but we have
   no information about the remaining eigenvalues and eigenfunctions.

\item
   We say that a system is (completely)-integrable if it admits $n$
   algebraically independent partial differential operators with variable coefficients $I_0={\cal H}, I_1,\cdots, I_{n-1}$ such that $[I_i,I_j]=0$ for $0\le i,j\le n-1$.
   Here, $[A,B]=AB-BA$ is the commutator. The system is {\bf superintegrable} if there are $s \ge 1$ additional partial differential operators with variable coefficients $I_n,I_{n+1},\cdots,I_{n+s-1}$ such that $[{\cal H},I_j]=0$ for $0\le j\le n+s-1$ and the set $\{I_j:\ 0\le j\le n+s-1\}$ is algebraically independent. If $s=n-1$, apparently the maximum possible, the system is {\bf maximally superintegrable}.
   The operator $I_j$ is called the integral or the symmetry. The integrals are chosen to be of a minimal order, in this case they are called {\bf basis} integrals. The {\bf order} of a (super)integrable system is the maximum order of the basis integrals (symmetries) $\{I_\ell:\ \ell=1,2,\cdots,n+s-1\}$. The basis integrals  $I_j$ of a superintegrable system generate a non-abelian algebra under the Lie commutator, not usually a Lie algebra. The structure and representation theory of this {\bf algebra of integrals} provides information about the spectral decomposition of the quantum system. In particular, maximal superintegrability captures the properties of quantum Hamiltonian systems that allow the Schr\"odinger eigenvalue problem $H\Psi=E\Psi$ to be solved exactly, analytically and algebraically.

\item
   Let ${\mathbf U_h}$ be an algebra of differential operators  that is  finitely-generated by ${\mathbf h}$. If ${\mathbf h}$ is a Lie algebra, ${\mathbf U_h}$ is its universal enveloping algebra. We say that a quantum system has {\bf hidden} algebra ${\mathbf U_h}$ if the Hamiltonian $I_0={\cal H}$ is an element of ${\mathbf U_h}$. In all so far known examples of this,  not only the Hamiltonian but all integrals are elements of ${\mathbf U_h}$. In this case the algebra of integrals is a sub-algebra of the hidden algebra. A trivial example of  hidden algebra is the Heisenberg-Weyl algebra. The first non-trivial example is  ${\mathbf h} =sl_2$ realized by the first order differential operators on $RP^1$. It is the explanation for the (quasi)-exact-solvability of many one-dimensional Schr\"odinger operators. In general, if the hidden algebra ${\mathbf U_h}$ has a finite-dimensional representation, the Hamiltonian (and sometimes integrals) has a finite-dimensional invariant subspace which coincides with finite-dimensional representation space of the hidden algebra. The Schr\"odinger eigenvalue problem ${\cal H}\Psi=E\Psi$ can be solved by algebraic means for elements of
   the finite-dimensional representation space.

\end{itemize}

Quantum systems and their classical analogs that can be solved exactly have been  of enormous historical importance: the harmonic oscillator,
the Kepler system (and the Hohmann transfer, used in celestial navigation), the quantum 2-body Coulomb system and, in particular, hydrogen atom 
(and its use to develop a perturbation theory for the periodic table of the elements), etc. The discovery and analysis of such systems is clearly of importance. There are close relations between the four types of systems listed previously, relations that are not yet well understood. For example, in \cite{TTW2001} there is a conjecture that all 2nd order superintegrable systems in Euclidean space $E^n$ are exactly solvable.
In this paper we shed more light on this and related conjectures by exhibiting a family of QES systems on the $n$-sphere that are non-maximally superintegrable for $n\ge 3$ and never exactly solvable. (They admit $(2n-2)$ second order symmetry operators (integrals).) This family contracts to a family of QES systems on $E^n$ that are non-maximally superintegrable for $n \ge 3$.

\section{An exactly-solvable problem}

Many years ago it was shown that the Lauricella polynomials $F_A$ (see e.g. \cite{Appell:1926}) are eigenfunctions of the algebraic operator with polynomial coefficients \cite{KMT:1990},
\begin{equation}
\label{ham}
 h^{(ES)}\ =\ \sum_{i,j=1}^n (x_i \de_{ij} - x_i x_j) \pa_i \pa_j\ +\ \sum_{i=1}^n \bigg(\frac{1}{2}+\ga_i -
 (G+\frac{n+1}{2}) x_i \bigg)\pa_i
 \ ,
\end{equation}
where $\de_{ij}$ is the Kronecker symbol, $\ga_i, i=1,2\ldots n, (n+1)$  are parameters and $G=\sum_{\ell=1}^{n+1} \ga_\ell$. At $n=1$ the Lauricella polynomials become Jacobi polynomials, while at $n=2$ they appear in the Krall-Sheffer description of polynomial eigenfunctions of a certain 2D eigenvalue problem. Choosing
\begin{equation}
\label{psi0}
    \Psi_0\ =\ x_1^{\frac{\ga_1}{2}} x_2^{\frac{\ga_2}{2}} \ldots x_n^{\frac{\ga_n}{2}}\ (1 - x)^{\frac{\ga_{n+1}}{2}} \ ,
\end{equation}
where $x = \sum_{i=1}^n x_i$, the operator (\ref{ham}) can be gauge rotated to the Schr\"odinger operator,
\begin{equation}
\label{Ham}
{\cal H}^{(ES)}\ = \ - \Psi_0 h^{(ES)} {\Psi_0}^{-1}\ \equiv \ -\De_g \ +\ \frac{1}{4} (V_0 - E_0)\ ,
\end{equation}
which is evidently Hermitian. Hence the Lauricella polynomials are orthogonal w.r.t. weight factor
$\Psi_0^2$.
Here $\De_g$ is the Laplace-Beltrami operator with contravariant metric
\begin{equation}
\label{gij}
    g^{ij}\ =\ x_i \de_{ij} - x_i x_j\ ,
\end{equation}
its determinant, $\det g^{ij} = g^{-1}$, where $g=\det g_{ij}$,
\begin{equation}
\label{g}
      g^{-1}\ =\ x_1 x_2 \ldots x_n (1-x)\ ,\quad x = \sum_{i=1}^n x_i\ ,
\end{equation}
(cf. (\ref{psi0})) and
\[
   g_{ij}\ =\ \frac{1}{1-x} - \frac{\de_{ij}}{x_i}\ .
\]
It can be found that scalar curvature to this metric is constant.

The potential in (\ref{Ham}) has the form
\begin{equation}
\label{potential}
 V_0\ =\ \sum_{i=1}^n \frac{\ga_i (\ga_i-1)}{x_i} + \frac{\ga_{n+1} (\ga_{n+1}-1)}{1 - x}\ ,
\end{equation}
where $\Psi_0$ (\ref{psi0}) plays a role of the ground state eigenfunction and
\[
       E_0\ =\ [G^2 + (n-1) G + 1]\ ,
\]
is the energy of ground state. Boundaries of the configuration space (domain) for (\ref{Ham}) are determined by zeros of $\Psi_0$. It defines the domain as simplex (pyramid with a regular main face (base)),
\[
   x_1 \geq 0\ ,\ x_2 \geq 0\ ,\ \ldots \ x_n \geq 0\ ,\ 1 \geq x \geq 0\ .
\]

One can exhibit a basis for the $\frac{n(n+1)}{2}$-dimensional space  of  second order differential operators which commute with $h^{(ES)}$ (\ref{ham}):
\begin{equation}\label{I_{ij}}
  I_{ij}\ = \ x_i x_j (\pa_i - \pa_j)^2 + [(\ga_i x_j - \ga_j x_i)+\half (x_j - x_i)]
  (\pa_i - \pa_j)\ ,\ 1 \leq i < j \leq n\ ,
\end{equation}
\begin{equation}\label{I_{i}}
I_{i}\ = \ x_i (1-x) \pa_i^2 + \big(\ga_i (1-x) - \ga_{n+1} x_i +\half ((1-x) +(2n+1) x_i)
\big) \pa_i \ ,\ 1 \leq i \leq n\ .
\end{equation}
All these operators are linearly independent while any $2n$ subset of them is algebraically dependent. The system is maximally 2nd order superintegrable. We note that the Hamiltonian (\ref{ham}) belongs to the space of integrals
\[
h^{(ES)}=\sum_{1\le i<j\le n}I_{ij}+\sum_{1\le i\le n}I_i\ .
\]

To identify the Riemannian space we introduce Cartesian coordinates $s_1, s_2, \ldots s_{0}$
in $(n+1)$-dimensional Euclidean space and assume that
\[
   x_1=s_1^2\ ,\ x_2=s_2^2\ ,\ \ldots \ x_n=s_n^2\ ,
\]
\begin{equation}
\label{s-coord}
    1 - x\ =\ {s_0}^2 \ ,
\end{equation}
which implies that the restriction $\sum_0^{n} s_i^2 =1$ (see \cite{KMT:1990} and references therein). Thus, $x = \sum_1^n s_i^2$ has the meaning of an $n$-dimensional section of the $n$-sphere, $s_0 = \mbox{const}$. Defining the element of the distance as
\[
     ds^2\ =\ \sum_0^n ds_i^2\ ,
\]
we find that in $x$-coordinates
\[
      ds^2\ =\ \frac{1}{4}\sum_{i,j=0}^n \ \left(\frac{1}{1-x} - \frac{\de_{ij}}{x_i}\right)\ dx_i dx_j\ .
\]

Thus, the space corresponds to a portion of the $n$-sphere $S^n$. We can consider the
coordinates $\{x_i\}$ for $x_i \geq 0$ and $1 \geq x \geq 0$ as covering the portion of the $n$-sphere given by $s_i \geq 0$ at $\sum_{i=0}^n s_i^2 = 1$. The metric of the constant curvature
in Cartesian coordinates reads
\begin{equation}
\label{gij-s}
    4 g^{ij}\ =\ \de_{ij} - s_i s_j\ .
\end{equation}

It is worth noting the explicit form of the $(n+1)$-parametric exactly-solvable potential in Cartesian coordinates
\begin{equation}
\label{potential-s}
 V_0\ =\ \sum_{i=1}^n \frac{\ga_i (\ga_i-1)}{s_i^2} + \frac{\ga_{n+1} (\ga_{n+1}-1)}{1 - s^2}\ ,
\end{equation}
where $s^2=\sum_1^n s_i^2$. Hence, the discrete symmetry of the Hamiltonian (\ref{Ham}) in Cartesian coordinates is ${\bf Z_2}^{\oplus n} \otimes S_n$ (reflections plus permutation), which is the symmetry of the Weyl group $BC_n$.

\section{The QES construction}
Let us consider the algebra $gl_n$ realized by the first order differential operators (see e.g. R\"uhl-Turbiner \cite{RT:1995})

\begin{eqnarray}
 {\cal J}_i^- &=& \frac{\pa}{\pa x_i}\ ,\qquad \quad
i=1,2, \ldots n , \non \\
 {{\cal J}_{ij}}^0 &=&
    x_i \frac{\pa}{\pa x_j}\ , \qquad i,j=1,2, \ldots n \ ,
 \non \\
{\cal J}^0(k) &=& \sum_{i=1}^{n} x_i\frac{\pa}{\pa x_i} - k\, ,
 \non \\
 {\cal J}_i^+(k) &=& x_i {\cal J}^0(k)\ =\
    x_i\, \left( \sum_{j=1}^{2} x_j\frac{\pa}{\pa x_j} - k \right)\ ,
       \quad i=1,2, \ldots n
\label{gln}
\end{eqnarray}
where $k$ is parameter. It is evident that the operator $h^{(ES)}$ can be rewritten in terms of generators ${\cal J}_i^-, {{\cal J}_{ij}}^0$,
\begin{equation}
\label{hamJ}
 h^{(ES)}\ =\ \sum_{i,j=1}^n (\de_{ij}{{\cal J}_{ii}}^0 {\cal J}_j^- - {{\cal J}_{ii}}^0{{\cal J}_{jj}}^0) \ +\ \sum_{i=1}^n \bigg((\frac{1}{2}+\ga_i){\cal J}_i^- -
 (G+\frac{n+1}{2}) {{\cal J}_{ii}}^0 \bigg)
 \ ,
\end{equation}
which span maximal affine subalgebra of $gl_n$. $h^{(ES)}$ has infinitely many finite-dimensional invariant subspaces
\begin{equation}
\label{Pk}
 {\cal P}_k^{(n)}\ =\ \langle {x_1}^{p_1}
 {x_2}^{p_2}\ldots
 {x_n}^{p_{n}}
 \vert \ 0 \le \Sigma p_i \le k \rangle\ ,
\end{equation}
when $k=0,1,\ldots$\,, which form flag:
\[
{\cal P}_0^{(n)} \subset  {\cal P}_1^{(n)} \subset {\cal P}_2^{(n)} \subset \ldots
 \subset  {\cal P}_k^{(n)}  \subset \ldots {\cal P}\ .
\]
Thus, the operator $h^{(ES)}$ is exactly-solvable \cite{Turbiner:1994}.
This  allows us  to find the spectra of (\ref{ham})
\begin{equation}
\label{ham-E}
 \ep_k\ =\ -k (k + G + \frac{n-1}{2})\ .
\end{equation}
The spectral  degeneracy is determined by the number of partitions of $k$ to sum of $n$ integer numbers (including zeros), the same as for $n$-dimensional harmonic oscillator.
Hence, the Lauricella polynomial is nothing but the element of the representation space of the algebra $gl_n$ realized as (\ref{gln}).
The spectra of exactly solvable Hamiltonian (\ref{Ham}),
$E_k^{(n)} = E_0 - \ep_k$, is equal to
\begin{equation}
\label{Ham-E}
 E_k^{(n)}\ =\ k (k + G +\frac{n-1}{2}) + [G^2 + (n-1) G + 1]\ ,
\end{equation}
which depends on a single parameter $G$. It is quadratic in quantum number $k$, which is typical for exact-solvable trigonometric (Sutherland) models in flat space. Similarly to the operator $h^{(ES)}$ the integrals $I_{ij}, I_{i}$ can be rewritten in $gl_n$ algebra generators.

The exactly-solvable operator (\ref{ham}) can be easily generalized to the quasi-exactly-solvable one by adding the sum of all raising generators (\ref{gln}),
\begin{equation}
\label{ham-qes}
   h^{(QES)}\ =\ h^{(ES)}\ + a \sum_{i=1}^n {\cal J}_i^+(k) \ ,
\end{equation}
with parameter $a$. Now this operator has a single invariant subspace (\ref{Pk}) in $n$-variate polynomials. The $n-1$ operators
\[
L_1=I_{12}\ ,\ L_2=I_{13}+I_{23}\ ,\ \cdots\ ,\ L_{n-1}=I_{1n}+I_{2n}+\cdots +I_{n-1\ n}\ ,
\]
span commutative algebra and leave the subspace ${\cal P}_k^{(n)}$ invariant. Thus, there exists an eigenpolynomial $\psi \in {\cal P}_k^{(n)}$ which is common for all $L$ operators and $h^{(QES)}$ such that
\begin{equation}
\label{c-eigens}
    L_1\Psi=c_1\psi\ ,\ (L_j + c_{j-1})\psi=c_j\psi\ ,\ j=2,\cdots,n-1\ ,\ h^{(QES)}\psi=E\psi\ ,
\end{equation}
where $c_i, i=1,\ldots, n-1$, \footnote{they have a meaning of separation constants, see below (\ref{ordsepeqns})} and $E$ are eigenvalues.
The total number of such eigenpolynomials is equal to $$\dim {\cal P}_k^{(n)} =\sum_{j=1}^k \frac{(n)_j}{j!}\ .$$

The common eigenfunctions form a basis for the subspace. They determine separation of variables for the eigenfunctions in spherical coordinates.

\subsection{2nd order operators commuting with  $h^{(QES)}$}

For $n\ge 2$, the operators $I_i$ do not commute with $h^{(QES)}$,
but the $n(n-1)/2$ linearly independent operators $I_{ij}$ do commute with this Hamiltonian.
Thus, the dimension of the space of 2nd order symmetries is $(n^2-n+2)/2$. For maximal superintegrability we must have $2n-1$ algebraically independent symmetries. We can predict that for $n=2$ the system will be merely integrable and we expect no degenerate energy eigenvalues.
For $n=3$, there are 4 algebraically independent symmetries, so the system is non-maximally superintegrable. For $n=4$ there are 7 symmetries, but we will see that only 6 are algebraically independent so the system is again
nonmaximally superintegrable, as is the case for all $n\ge 4$. For $n>4$ any $2n-1$ subset of symmetries is algebraically dependent.

For $n\ge 3$ the symmetries generate a noncommutative algebra by taking commutators, so there will be degenerate spectra.
For the lowest dimensional case of this, $n=3$, the algebra generated by $L_{12},L_{13},L_{23}$ is quadratic.
There is a single commutator
\[ R=[L_{12},L_{13}]=[L_{13},L_{23}]=[L_{12},L_{23}].\]
The structure equations are
\[
  [L_{ij},R]=4\ep_{ijk}\left(
  \{L_{ij},L_{ik}-L_{jk}\}+2(1+2a_j)L_{ik}-2(1+2a_i)L_{jk}+2(a_i-a_j)
  \right)\ ,
\]
where $a_i=\ga_i(\ga_i-1)$,  $ \{A,B\}=AB+BA$ is the anticommutator,
\ $i,j,k$ are pairwise distinct integers $i\le i,j,k \le 3$ and $\ep_{ijk}$ is the completely skew-symmetric tensor such that $\ep_{123}=1$. The Casimir operator is
\[
 R^2\ =\ \frac83\{ L_{12},L_{13},L_{23}\}-4(3+4a_3)L_{12}^2-4(3+4a_1)L_{23}^2-4(3+4a_2)L_{13}^2\ +
\]
\[
 \frac{52}{3}(\{L_{12},L_{13}+L_{23}\}
 +\{L_{13},L_{23}\})+\frac{16}{3}(1 + 11a_3)L_{12}+\frac{16}{3}(1 + 11a_1)L_{23}
\]
\[
 + \frac{16}{3}(1 + 11 a_2)L_{31}
 + 64a_1 a_2 a_3 + 48(a_1 a_2+a_2 a_3+a_3 a_1)+\frac{32}{3}(a_1+a_2+a_3)\ ,
\]
 where
\[
\{ A,B,C  \} =ABC +ACB+BAC+BCA+CAB+CBA\
\]
is a symmetrizer. This is exactly the symmetry algebra for the generic system on the 2-sphere, called $S9$ in the list in \cite{KKMP}. The irreducible representations of physical importance have been worked out in \cite{KMP2007a}. From these results we see that the spectra of the $L_{ij}$ operators and the multiplicities of the energy spectra can be computed algebraically, but not the energy spectrum itself. For $n=4$ the symmetry algebra generated by the $L_{ij}$ symmetries is isomorphic that of the generic system on the 3-sphere; its structure is determined in \cite{KMP2011}.

\subsection{The separation equations}

The eigenfunctions $\Psi(x)$ of Hamiltonian $ h^{(QES)}$, ($h^{(QES)}\Psi=E\Psi$), are separable in the spherical coordinates $\{u_i\}$ where
\[
x_1=u_1 u_2\cdots u_n\ ,\ x_2=(1-u_1)u_2\cdots u_n\ ,\ \cdots,\  x_{n-1}=(1-u_{n-2})u_{n-1}u_n\ ,
\]
\begin{equation}
\label{sphericalcoords}
x_n=(1-u_{n-1})u_n\ ,\ x=u_n\ .
\end{equation}
In terms of angles, one usually writes $u_i=\sin^2\phi_i$. Taking
\footnote{This implies a quite non-trivial factorization: a polynomial in $x_i$ becomes a product of polynomials in $u_j$;
it also implies that $$x_1^{q_1} x_2^{q_2} \ldots x_n^{q_n} =\left(\prod_{j=1}^{n-1}(1-u_j)^{q_{j+1}}\right)\left(\prod_{i=1}^n u_i^{m_i}\right) \ ,$$ where $m_s=\sum_{j=1}^s q_j$},
\begin{equation}
\label{factorization}
\Psi(x(u))\ =\ \prod_{i=1}^n U_i(u_i)\ ,
\end{equation}
we obtain separation equations
\[
u_1(1-u_1)\frac{d^2U_1}{du_1^2}+\left(G_1+\frac12-u_1(1+G_2)\right)
\frac{dU_1}{du_1}-c_1U_1=0\ ,
\]
\[
u_{2}(1-u_{2})\frac {d^2U_{2}}{du_{2}^2}+\left(G_2+\frac{2}{2}-u_{2}(\frac{3}{2}
+G_3)\right)\frac{dU_{2}}{du_{2}}+\left(\frac{c_1}{u_{2}}-c_2\right)U_{2}=0\ ,
\]
\begin{equation}
\label{ordsepeqns}
\hfill \cdots \hfill
\end{equation}
\[
 u_{\ell}(1-u_{\ell})\frac{d^2U_{\ell}}{du_{\ell}^2}+\left(G_{\ell} +
 \frac{\ell}{2}-u_{\ell}(\frac{\ell+1}{2} +G_{\ell+1})\right)\frac{dU_{\ell}}{du_{\ell}}+\left(\frac{c_{\ell-1}}{u_{\ell}}- c_{\ell}\right)U_{\ell}=0\ ,
\]
\[
    \hfill \cdots \hfill
\]
\[
 u_{n-1}(1-u_{n-1})\frac{d^2U_{n-1}}{du_{n-1}^2}+\left(G_{n-1} +
 \frac{n-1}{2}-u_{n-1}(\frac{n}{2} +G_n)\right)\frac{dU_{n-1}}{du_{n-1}}+\left(\frac{c_{n-2}}{u_{n-1}}- c_{n-1}\right)U_{n-1}=0\ ,
 \]
and
\begin{equation}
\label{specsepeqn}
u_n(1-u_n)\frac{d^2U_n}{du_n^2}+\left(G_n + \frac{n}{2}-u_{n}(-\frac{n+1}{2}
 + G )+ a u_n^2\right)
\frac{dU_{n}}{du_{n}}+\left(\frac{c_{n-1}}{u_{n}}- a k u_n - E\right)U_{n}=0\ .
\end{equation}
Here
\[
    G_j\ =\ \sum_{i=1}^j \ga_i\ ,\ G_{n+1} = G\ .
\]
Any equation (\ref{ordsepeqns}) as well as (\ref{specsepeqn}) can be written in a form of eigenvalue problem
\[
     L_{\ell} U_\ell \ =\ c_{\ell} U_\ell\ , \ \ell=1,\cdots,n\ ,
\]
where $c_{\ell}$ plays a role of spectral parameter. All equations (\ref{ordsepeqns}), (\ref{specsepeqn}) together can be considered as $n$-spectral (multi-spectral) problem.
It can be shown that by a gauge rotation the operators $L_{\ell}, \ell=2,\cdots,(n-1)$ can be reduced to the hypergeometric operator,
\begin{equation}
\label{ordsepeqns-H}
     u_{\ell}^{-A_{\ell}} L_{\ell} u_{\ell}^{A_{\ell}}\ =\ u_{\ell}(1-u_{\ell})\frac{d^2}{du_{\ell}^2}+
     \left(2A_{\ell} + G_{\ell} + \frac{\ell}{2} -
     u_{\ell}(\frac{\ell+1}{2} + G_{\ell+1}+2A_{\ell})\right)\frac{d}{du_{\ell}}\ ,
\end{equation}
where $A_1=0$ and $A_{\ell}$ should be a solution of the equation
\[
    A_{\ell}^2 + A_{\ell} (G_{\ell} + \frac{\ell}{2}-1)+ c_{\ell-1}\ =\ 0\ .
\]
The spectrum of the  hypergeometric operator is quadratic. We find
\begin{equation}
\label{eigen-confH} A_\ell=\sum_{i=1}^{\ell-1} q_i,\quad
    c_{\ell} = -A_{\ell+1}(A_{\ell+1}+G_{\ell+1}+\frac{\ell-1}{2}) ,\
\end{equation}
where  the eigenfunctions are hypergeometric polynomials of degree $q_\ell$.

In a similar way the operator $L_n$ is reduced to the confluent Heun operator
\begin{equation}
\label{specsepeqn-H}
     u_{n}^{-A_{n}}\ L_{n}\ u_{n}^{A_{n}}\ =\
\end{equation}
\[
     u_{n}(1-u_{n})\frac{d^2}{du_{n}^2}+\left(2A_{n} + G_{n} +
 \frac{n}{2}-u_{n}(-\frac{n+1}{2} + G + 2A_{n})+ a u_n^2 \right)\frac{d}{du_{n}} - a (k-A_n) u_n\ ,
\]
It turns out that $(k - A_n) = m$ must be a nonnegative integer! It defines the degree $m$ of $(m+1)$ polynomial eigenfunctions of the confluent Heun operator (\ref{specsepeqn-H}).

Above analysis implies a certain modification of (\ref{factorization})
\begin{equation}
\label{factorization-M}
\Psi(x(u))\ =\ u_2^{A_2}\ldots u_n^{A_n}\prod_{i=1}^n V_i(u_i)\ .
\end{equation}
where  all $V_i(u_i)$ are polynomials if $\Psi(x)$ is a polynomial eigenfunction of $ h^{(QES)}$.
 Eigenvalues $c_\ell,\  \ell=1,\cdots,n-1$ play a role of separation constants. The solutions to the modified separation equations are
\[
V_\ell(u_\ell)={}_2F_1\left(\begin{array}{cc} -q_\ell,& 2\sum_{i=1}^{\ell-1}q_i+q_\ell
+G_{\ell+1} + \frac{\ell-1}{2}\\ 2\sum_{i=1}^{\ell-1}q_i+G_{\ell} + \frac{\ell}{2}\end{array}; u_\ell\right)\ ,
\]
for $\ell=2,\cdots, n-1$.  The function $V_n(u_n)$ (up to a factor) is a polynomial of order $m$, being a confluent Heun polynomial.

\subsection{A (quasi)-exactly-solvable problem in flat space: a connection}

All separation equations operators (\ref{ordsepeqns-H}), (\ref{specsepeqn-H}) have an interesting common property: a trigonometric change of variables
\[
         u_{\ell} = \sin^2 y_{\ell}
\]
converts the $1D$ Laplace-Beltrami operator to the $1D$ Laplace operator (in the second derivative terms). Thus, by
changing variables and making appropriate gauge rotations we convert the operators (\ref{ordsepeqns-H}), (\ref{specsepeqn-H}) into the $1D$ Schr\"odinger operators
with modified P\"oschl-Teller potential for the case of (\ref{ordsepeqns-H})
and to the quasi-exactly-solvable modified P\"oschl-Teller potential for the case of (\ref{specsepeqn-H}) (see e.g. \cite{Turbiner:2013}).
Summing up all these equations we end up with an $n$-dimensional Schr\"odinger operator in flat space which is equivalent to the original $n$-dimensional Schr\"odinger operator
on the $n$-sphere.

\subsection{Solutions of the $n$-sphere equations for $n=1,2,3$, $k=0,1,2$}
\begin{itemize}
\item The three-sphere $S^3$
\begin{enumerate}
\item $n=3$, $k=2$: The invariant subspace is 10-dimensional. We choose a common eigenbasis of the commuting symmetry algebra operators
\begin{equation} \label{I12} I_{12} = \ x_1 x_2 (\pa_1 - \pa_2)^2 + [(\ga_1 x_2 - \ga_2 x_1)+\half (x_2 - x_1)]
  (\pa_1- \pa_2)\end{equation} and
\begin{equation} \label{I13+I23}  I_{13}+I_{23}=  \ x_1 x_3 (\pa_1 - \pa_2)^2 + [(\ga_1 x_3 - \ga_3x_1)+\half (x_3 - x_1)]
  (\pa_1- \pa_3)+ \end{equation}
\[ x_2 x_3 (\pa_2 - \pa_3)^2 + [(\ga_2 x_3 - \ga_3x_2)+\half (x_3 - x_2)]
  (\pa_2- \pa_3).\]
For  $q_1+q_2=2$ there is a single energy eigenvalue  $E=2(1-G)$ of multiplicity 3.
For $q_1+q_2=1$ there are two energy eigenvalues
\[
E_\pm=2-\frac{3G}{2} \pm \frac12\sqrt{G^2 - 2a (7 +2 G_3)}\ ,
\]
each eigenvalue of multiplicity 2.
For $q_1=q_2=0$ there are 3 energy eigenvalues, each of multiplicity 1. They satisfy the cubic equation
\[
   E^3 + (3 G - 4) E^2+2\left(2a(G_3 + 2)+(G-1)(G-2)\right)E
\]
\[+
 2a(2G_3+3)(G-1)=0
\]
 \item $n=3$, $k=1$: The invariant subspace is 4-dimensional.  We choose a common eigenbasis of the symmetries $I_{12}$ and $I_{13}+I_{23}$. For $q_1+q_2=1$ there is a single energy eigenvalue
 $E=2-G$ of multiplicity 2. For $q_1=q_2=0$ there are two energy eigenvalues
\[
     E_\pm=1-\frac{G}{2} \pm\frac12\sqrt{(G-2)^2- 2a(3 + 2G_3 )}\ ,
\]
 each of multiplicity 1.
 \item  $n=3$, $k=0$: We choose a common eigenbasis of the symmetries $I_{12}$ and $I_{13}+I_{23}$.
 The invariant eigenspace is 1-dimensional and the energy eigenvalue is $E=0$.\end{enumerate}

 \item The two-sphere  $S^2$

\begin{enumerate}
 \item $n=2$, $k=2$: The invariant subspace is 6-dimensional.  We choose an eigenbasis of the symmetry algebra
\begin{equation} \label{I12a} I_{12} = \ x_1 x_2 (\pa_1 - \pa_2)^2 + [(\ga_1 x_2 - \ga_2 x_1)+\half (x_2 - x_1)]
  (\pa_1- \pa_2).\end{equation} For $q_1=2$ there is a single energy eigenvalue
 $E=1-2G$ of multiplicity 1. For $q_1=1$ there are two energy eigenvalues
\[
   E_\pm\ =\ \frac54-\frac{3G}{2} \pm \frac12\sqrt{(G + \frac12)^2-4a(3 + G_2) }\ ,
\]
each eigenvalue of multiplicity 1. For $q_1=0$ there are 3 energy eigenvalues, each of multiplicity 1. They satisfy the cubic equation
\[
  2E^3+(6G - 5)E^2+\left(4a(3 + 2 G_2) +(1-2G)
  (3-2G)\right)E\]
\[
 +4a(G_2 + 1)(2G-1)=0.\]
 \item  $n=2$, $k=1$: The invariant subspace is 3-dimensional.  We choose an eigenbasis of the symmetry $I_{12}$. For $q_1=1$ there is a single energy eigenvalue
 $E=\frac32-G$ of multiplicity 1. For $q_1=0$ there are two energy eigenvalues
\[
E_\pm=\frac34-\frac{G}{2} \pm \frac12\sqrt{(G-\frac32)^2-4a(1+G_2)}\ ,
\]
each eigenvalue of multiplicity 1.
\item $n=2$, $k=0$:  We choose an eigenbasis of the symmetry $I_{12}$. The invariant eigenspace is 1-dimensional and the energy eigenvalue is $E=0$.
\end{enumerate}

\item The one-sphere  $S^1$

\begin{enumerate}
 \item $n=1$, $k=2$: The invariant subspace is 3-dimensional. There are 3 energy eigenvalues. They satisfy the cubic equation
\[
  E^3 + (3G + 5)E^2 + 2 \left(2a(G_1+1)+(G+2)(G+1)\right)E
 +2a(2G_1 + 1)(G+2)=0\ .
\]
Hence, the eigenvalues are branches of 3-valued analytic function in $a$. Ramification points are square-root singularities. The corresponding eigenfunctions of $h^{(QES)}$ have a form $x^2 + A x + B$.

\item $n=1$, $k=1$: The invariant subspace is 2-dimensional. There are two energy eigenvalues
\[
   E_\pm\ =\ -\frac{G}{2} -\half \pm \frac12\sqrt{(G + 1)^2 - 2a(1 +2 G_1)} \ ,
\]
They form double-sheeted Riemann surface. The corresponding eigenfunctions of $h^{(QES)}$ are
\[
   \phi_{\pm} = x + \frac{1 + 2 G_1}{2 E_{\pm}}
\]

\item $n=1$, $k=0$: The invariant subspace is one-dimensional. There is single energy eigenvalue, $E=0$.

\end{enumerate}

\end{itemize}

\subsection{The quantum Hamiltonian}
The operator (\ref{ham-qes}) can be gauge-rotated with
\[
      {\tilde \Psi}_0^{(QES)}\ =\ \exp \{ { - \frac{a}{2} \sum_1^n x_i} \}\ .
\]
It leads to a potential additional to the potential $V_0$,
\[
     a^2 (\sum_1^n x_i)^2 -  \bigg(a^2 - a (2 G + n + 1 - 4k)\bigg) (\sum_1^n x_i) \ .
\]
Eventually, we arrive at the QES potential

\begin{equation}
\label{potential-qes}
 V\ =\ a^2 x^2 - a (a - 2 G - n - 1 + 4k) x + \sum_{i=1}^n \frac{\ga_i (\ga_i-1)}{x_i} + \frac{\ga_{n+1} (\ga_{n+1}-1)}{1 - x}\ ,
\end{equation}
where the ``algebraic" eigenfunctions have the form

\[
     \Psi_k^{(QES)}\ =\ x_1^{\frac{\ga_1}{2}} x_2^{\frac{\ga_2}{2}} \ldots x_n^{\frac{\ga_n}{2}}\ (1 - x)^{\frac{\ga_{n+1}}{2}} P_{k,\ell} (x_1,\ldots, x_n) \exp \{ - \frac{a}{2} x \}\ ,\
     \ell = 1, \ldots \dim {\cal P}_k^{(n)}\ ,
\]
with $P_{k,\ell} \in {\cal P}_k^{(n)}$. In Cartesian coordinates the potential (\ref{potential-qes})
has the  form
\begin{equation}
\label{potential-qes-Car}
 V^{(QES)}\ =\ a^2 s^4 - a (a - 2 G - n - 1 + 4k) s^2 + \sum_{i=1}^n \frac{\ga_i (\ga_i-1)}{s_i^2} + \frac{\ga_{n+1} (\ga_{n+1}-1)}{1 - s^2}\ ,
\end{equation}
while the ``algebraic" eigenfunctions have the form

\[
     \Psi_k^{(QES)}\ =\ s_1^{\ga_1} s_2^{\ga_2} \ldots s_n^{\ga_n}\ (1 - s^2)^{\frac{\ga_{n+1}}{2}} P_{k,\ell} (s_1^2,\ldots, s_n^2) \exp \{ - \frac{a}{2} s^2 \}\ ,\
     \ell = 1, \ldots \dim {\cal P}_k^{(n)}\ .
\]
The eventual form of the quasi-exactly-solvable Hamiltonian can be obtained making a gauge rotation of (\ref{ham-qes}) written in Cartesian coordinates with gauge factor $\Psi_0^{(QES)}$,
\[
   {\cal H}^{(QES)}\ =\ -\De_g(s) + V^{(QES)}\ ,
\]
where the Laplace-Beltrami operator has metric (\ref{gij-s1}). Hence, its symmetry is ${\bf Z_2}^{\oplus n} \otimes S_n$, which is the symmetry of the Weyl group $BC_n$.

For general $n$ the eigenvalue equation for ${\cal H}^{(QES)}$ is separable in many coordinate systems $\{v_1,\cdots,v_{n-1},u_n\}$, not just the spherical coordinate system treated here. All of these systems take the form
\[
x_1=X_1({\bf v})u_n\ ,\  \cdots, \  x_{n-1}=X_{n-1}({\bf v})u_n\ ,\ x_n= X_n({\bf v})u_n\ ,
\]
where  $v_1,\cdots,v_{n-1}$   are any separable coordinates on the $(n-1)$-sphere, These can be polyspherical coordinates, ellipsoidal coordinates or mixtures of the two, as classified in \cite{KMJ,ERNIE}. The separation equations
(\ref{ordsepeqns}) are  replaced by new separation equations, some with hypergeometric polynomial solutions and some with Heun polynomial solutions.  However, the separation equation (\ref{specsepeqn}) is  common to all of them. Thus we see that this eigenvalue equation is exactly solvable in no separable coordinate system.

To make this clearer, we define new coordinates $z_1,\cdots, z_{n-1},r$ such that
\[
  x_\ell=rz_\ell,\quad x_n=r(1-z),\quad \ell=1,2,\cdots,n-1\ .
\]
Here $z=\sum_{\ell=1}^{n-1}z_\ell$ and we note that $x=r$. Then we find
\begin{equation}\label{radialsplit}
   h^{(QES)}=r(r-1)\pa_{rr}-\left(G_n+\frac{n}{2}+r(\frac{n+1}{2}-G)+ar^2\right)\pa_r +a k r +\frac{1}{r}h_{n-1}^{(ES)}\ ,
\end{equation}
where $h_{n-1}^{(ES)}$ is the exactly solvable Hamiltonian on the sphere $S^{n-1}$ expressed
in terms of the coordinates $z_\ell$. In general terms, it corresponds to the decomposition
$S^n \sim S^1 \times S^{n-1}$, realizing a separation of variable $r$ from a set of variables
which parametrize the sphere $S^{n-1}$. {\it This  is analogous to the decomposition
$E^n\sim R^+ \times S^{n-1}$ obtained by introducing spherical coordinates in Euclidean space,
for which $r$ is the radial coordinate. Note the feature of the sphere that the above decomposition can be recurrent: $S^n \sim \underbrace{S^1 \times \ldots S^1}_p \times S^{n-p}$. }

The $S^{n-1}$ piece of the decomposition contributes $2(n-1)-1$ algebraically independent 2nd order integrals and $n(n-1)/2$ linearly independent 2nd order integrals (as we have shown ({\ref{I_{ij}}), ({\ref{I_{i}})), and the $S^1$ piece contributes 1, (namely the Hamiltonian)  which gives $2n-2$ algebraically independent and $n(n-1)/2 + 1$ linearly independent integrals in total.  All these potentials have $(n+2)$ parameters.  For $n\ge 3$ the system is nonmaximally  superintegrable, one step below maximum superintegrable.

From this analysis we see that the symmetry algebra of $H^{(QES)}$ decomposes as the direct sum of  $H^{(QES)}$ itself and the symmetry algebra of the generic 2nd order superintegrable system on  $S^{n-1}$. The structure algebra is thus the sum of the structure algebra of   generic superintegrable system on  $S^{n-1}$ and a one-dimensional term generated by $H^{(QES)}$, which is in the center. The irreducible representations of this structure algebra are essentially those of the $S^{n-1}$ structure algebra. They allow us to determine algebraically the spectra of the
generators $L_{ij}$ and the multiplicities of the energy eigenvalues, but they  give no information about the values of the energy eigenvalues.

\section{A superintegrable QES system in $E^n$}\label{EuclideanQES}

We consider the Euclidean (flat space) Hamiltonian in Cartesian coordinates

\begin{equation}
H^{(ES)} \ = \ -\De_n +  \om^2\,(\sum_{j=1}^ny_j^2) + \sum_{j=1}^n \frac{{\ga'}_j^2-\frac{1}{4}}{y_j^2}\ ,
\label{H1}
\end{equation}
where $\De_n = \sum^n \pa_{y_i}^2$ is the Laplacian and $\om, \ga'$ are parameters.
This system described by (\ref{H1}) is well known to be maximally superintegrable.
Introducing the coordinates $Y_j=y_j^2$ in (\ref{H1}) we obtain
\begin{equation}
\label{H2}
H^{(ES)}=-2\sum_{j=1}^n\left(2 Y_j\pa^2 _{Y_j} + \pa_{Y_j}\right)+ \om^2\,\sum_{j=1}^nY_j
 + \sum_{j=1}^n \frac{{\ga'}_j^2-\frac{1}{4}}{Y_j} \ ,
\end{equation}
where the first term is the Laplace-Beltrami operator with flat metric $g^{ij}=4Y_i \de^{ij}$.
The ground state wave function takes the form
\[
  \psi_0 \ =\ e^{-\frac{\om}{2}\,\sum_{j=1}^n Y_j}\  \prod_{j=1}^nY_j^{\frac{1}{4}- \frac{\gamma'_j}{2}}  \ .
\]
Subtracting the ground state energy $E_0 = 2\om( \sum_{j=1}^n\ga'_j-n)$ from (\ref{H1}) and making
the gauge transformation of (\ref{H2}) with the gauge factor $\psi_0$ we obtain the
gauge rotated Hamiltonian
\begin{equation}
 {\hat h}^{(ES)}\ \equiv  \
 \psi_0^{-1}\,(H^{(ES)}-E_0)\,\psi_0   \ = \
 -4\, \left[\sum_{j=1}^n Y_j\,\pa^2_{Y_j} - \sum_{j=1}^n(\ga'_j-1
 +\om Y_j)\pa_{Y_j}\right ]\,,
\label{h2}
\end{equation}
(cf.(\ref{ham})), which is a sum of Hermite operators, and thus maps polynomials into polynomials
without increasing the overall degree. It is easy to check that (\ref{h2}) can be rewritten
in terms of $gl_n$-generators (\ref{gln}), ${\cal J}_i, {\cal J}^0_{ij}$ (where $x_i \rar Y_i$).
Thus, the model (\ref{H1}, (\ref{H2}) is $gl_n$ Lie-algebraic like the model (\ref{Ham}).

Now, we take the operator
\[
 B \equiv \  \,(\sum_{j=1}^n Y_j)\,(\sum_{j=1}^nY_j\,\pa_{Y_j} -k) = \
 \sum_{j=1}^n {\cal J}_j^+(k)\ ,
\]
where  $k$ is a nonnegative integer, and form the Hamiltonian
\[
{\hat h}^{(QES)} \ =\  {\hat h}^{(ES)} +  b B \ ,
\]
where $b$ is a parameter, cf.(\ref{ham-qes}).
The operator $B$ corresponds to additional terms in the Hamiltonian (\ref{H1}). In particular, it leads to first derivative terms. Using the gauge factor
\[
U\ =\ e^{\frac{b}{16}\ {\sum_{j=1}^n Y_j}^2} \,,
\]
we can get rid of all such first derivative terms in $\sim b$. Finally, the resulting Hamiltonian $ H^{(QES)} = U^{-1}\,\psi_0\,h^{(QES)}\, \psi_0^{-1}\,U$ reads

\begin{equation}
\label{HQES-E}
 H^{(QES)} \ =\ H^{(ES)} + \frac{b^2}{16}\,(\sum_{j=1}^nY_j)^3+\frac{b}{2}
 \left[ (\sum_{j=1}^nY_j)(\sum_{j=1}^n\gamma'_j-n-2k-1)
              +\om (\sum_{j=1}^nY_j)^2\right]\ ,
\end{equation}
which for $n=1$ corresponds to a celebrated quasi-exactly-solvable sextic potential \cite{Turbiner:1988}.
This system is closely related to our construction on spheres. Indeed, we can introduce new coordinates $R, Z_i$ by
\[
Y_i=RZ_i,\quad Y_n=R(1-Z),\ Z=\sum_{\ell=1}^{n-1}Z_\ell,\quad i=1,\cdots,n-1\ .
\]
Then
\begin{equation}
\label{radialsplite}
{\hat h}^{(QES)}(R)\ =\
 -4R\,\pa_R^2+\left(bR^2 + 4 \om R + 4\sum_{j=1}^n\ga'_j-4n\right)\pa_R-bk\,R+\frac{4}{R} {\hat h}^{(ES)}_{S^{n-1}}\ .
\end{equation}
If we set $\gamma'_j=-\gamma_j+\frac12,\ j=1,\cdots,n-1$ then  $h^{(ES)}_{S^{n-1}}$ is identical to the exactly solvable system (\ref{ham}), but on the $(n-1)$-sphere with coordinates $Z_i$.
The eigenvalue equation $ H^{(QES)}\Psi=E\Psi$ is separable in multiple coordinate systems, but all are of the form $R, U_i$ where the $U_i$ are separable coordinates  for the
equation $H^{(ES)}_{S^{n-1}}\Theta=\lambda\Theta$. The separation equations for the $U_i$ may or may not have hypergeometric solutions. However
the separation operator  for
$R$ takes the form
\[ -4R\,\partial_R^2+\left(bR^2+4\omega R+4\sum_{j=1}^n\gamma'_j-4n\right)\partial_R-bk\,R+\frac{4}{R}\lambda'_m,\]
which does not have hypergeometric solutions. There are polynomial solutions of order $k$. For these, $m=0,1,\cdots,k$. There are no polynomial solutions of order $>k$.
This system admits $2n-2$ algebraically independent 2nd order  integrals, so it is superintegrable for $n\ge 3$, one step below maximal
superintegrable. A key observation is that
this system  splits the space as $E^n\approx R^+ \times S^{n-1}$. The $S^{n-1}$ piece
contributes $2(n-1)-1$ algebraically independent 2nd order  integrals and $n(n-1)/2$ linearly independent 2nd order integrals
(as we have shown ({\ref{I_{ij}}), ({\ref{I_{i}})), and
the $R^+$ piece contributes 1, (namely the Hamiltonian)  which gives 2n-2 algebraically independent and $n(n-1)/2 + 1$ linearly
independent integrals in total.  All these potentials have $(n+2)$ parameters.

\section{Construction of QES systems via contractions}

By taking a series of contractions from the system $H^{(QES)}$ on the $n$-sphere
we can construct other QES systems on the $n$-sphere and $n$-dimensional Euclidean space, with a variety of potentials. Details about contractions and their relation to superintegrable systems, separation of variables and special functions can be found in many places; particularly relevant are \cite{Wigner, IPSW1996, KMP1999, KMP2013,KM2014}. Here we describe, briefly, how the Euclidean system of Section \ref{EuclideanQES}  arises as a contraction of the system on the $n$-sphere.

The operators
\[
J_{\ell m}=s_\ell\partial_m-s_m\partial_\ell,\  J_{\ell m}=-J_{m\ell},\ \ell\ne m, \quad \sum_0^n s_\ell^2 =1\ ,
\]
form a basis of the symmetry algebra $so(n+1)$ of $S^n$. Given a parameter $\epsilon\ne 0$ we define a new basis for  $so(n+1)$ by
\[
J'_{ij}=J_{ij},\ \quad P_j=\epsilon J_{0j},\quad  1\le i,j\le n\ .
\]
We can write the structure equations for $so(n+1)$, in terms of the new basis $\{ J'_{ij}, P_j\}$. In the limit as $\epsilon\to 0$ the change of basis becomes singular but the structure constants
go to a finite limit. Indeed, in the limit as $\epsilon\to 0$ we find that $\{ J'_{ij}, P_j\}$ satisfy the commutation relations for the Euclidean Lie algebra $e(n)$. This abstract Lie algebra contraction of $so(n+1)$ to $e(n)$ is implemented by the coordinate substitution $s_i=\epsilon x_i$, $1\le i\le n$, where the $x_i$ are Cartesian coordinates for $E^n$. In term of the polynomial coordinates on $S^n$ and $E^n$ we have $z_i=Z_i,\ i=1,\cdots, n$ and $r=\epsilon^2R$. Then, setting
\[
\gamma'_j=-\gamma_j+\frac12,\  j=1,\cdots,n-1,\quad \gamma_{n+1}'=\frac{\omega}{\epsilon^2},\quad a=-\frac{b}{\epsilon^4}\ ,
\]
we see that the operators (\ref{radialsplit}) and (\ref{radialsplite}) are related by
\[
h_{E^n}^{(QES)}=\lim_{\epsilon\to 0}4\epsilon^2 \, h_{S^n}^{(QES)}\ .
\]
There is a hierarchy of such contractions, all based on the system $h_{S^n}^{(QES)}$.

\section{Conclusions and discussion}

We have constructed  quasi-exactly solvable systems with $(n+2)$-parametric potential
on the $n$-sphere and on $n$-dimensional Euclidean space for every integer  $n$.
For $n=2$ the systems are merely integrable, but for $n \ge 3$ they are 2nd order  superintegrable, admitting $2(n-1)$ algebraically independent 2nd order integrals,
$1$ less than the maximal degree.
These systems are significant in several respects. One-dimensional QES systems have been constructed and related to superintegrable  systems on the $n$-sphere before, see \cite{KMP2006}, but those arose as separations equations for maximal superintegrable systems on the sphere that were exactly solvable. Here, the `parent' system on the sphere is not maximally superintegrable and is never exactly solvable. We have also shown how other QES
systems on constant curvature spaces can be obtained as contractions of the basic QES system on the sphere.

In the paper \cite{TTW2001} it was conjectured that all 2nd order superintegrable
systems in $n$-dimensional Euclidean space were exactly solvable. We know of no proof of this conjecture but there is a lot of evidence to support it. However, our examples show that a necessary condition for the validity of the conjecture is that the systems must be maximally superintegrable.

\section*{Acknowledgments}

  A.V.T. is thankful to University of Minnesota for kind hospitality extended to him where this work was initiated.
  The first author was partially supported by a grant from the Simons Foundation (\# 208754 to Willard Miller, Jr.).
  The second author is supported in part by the University Program FENOMEC, and by the PAPIIT
  grant {\bf IN109512} and CONACyT grant {\bf 166189}~(Mexico).

\newpage

\centerline{APPENDIX}

\begin{center}
    {\it Polynomial Metrics of the Laplace-Beltrami operator on the $n$-sphere $S^n$ in 
    invariant coordinates}
\end{center}

\bigskip

\begin{equation}
\label{LB}
    \De_g\ =\ \frac{1}{g^{1/2}}\frac{\pa}{\pa \ta_{a}}\ (g^{a b} g^{1/2})\ \frac{\pa}{\pa \ta_{b}}\ =\ g^{a b} \frac{\pa^2}{\pa \ta_{a}\pa \ta_{b}}\ +\ g^{b} \frac{\pa}{\pa \ta_{b}}\ ,\ g^{b} \equiv \frac{1}{g^{1/2}}\frac{\pa}{\pa \ta_{a}}\ (g^{a b} g^{1/2}) \ ,
\end{equation}

\bigskip

For $n$-sphere $S^n$ in Cartesian coordinates

\bigskip

\begin{equation}
\label{gij-s1}
    4 g^{ij}\ =\ \de_{ij} - s_i s_j\ ,
\end{equation}

\bigskip

$\bullet \quad n=1$

\bigskip

There exists a single discrete symmetry on the line: $Z_2 (s_1 \rar -s_1)$, hence  $\tau = x = s_1^2$ is invariant and

\[
        g^{11}\ =\ \ta (1-\ta)\ ,\ g^1 \pa_1 = (\half + \ta)\pa
\]

\bigskip

$\bullet \quad n=2$

\bigskip

The most general discrete symmetry
$$Z_2 (s_1 \lrar -s_1)\oplus Z_2 (s_1 \lrar -s_1) \oplus S_2 (s_1 \lrar s_2)\ ,$$ hence
$$\ta_1 = x + y = s_1^2 + s_2^2\ ,\ \ta_2 = x y = s_1^2 \ s_2^2$$ are invariants and

\[
        g^{11}\ =\ \ta_1 (1 - \ta_1)\ ,\ g^{22}\ =\ \ta_2 (\ta_1 - 4\ta_2)\ ,\
        g^{12}\ =\ g^{21}\ =\ 2\ta_2 (1 - \ta_1)\ ,
\]
\[
        g^{i} \pa_i\ =\ (1 - \frac{3}{2}\ta_1)\pa_1 + \half (\ta_1 - 10\ta_2)\pa_2 ,
\]

\end{document}